\documentclass[aps,prb,twocolumn,showpacs,showkeys,superscriptaddress,10pt]{revtex4-1}
\usepackage{amsmath}
\usepackage{graphics}

\include{MyMc}

\newcommand{\Cp}{\ensuremath{{J_c^+}}}
\newcommand{\Cm}{\ensuremath{{J_c^-}}}
\newcommand{\Cpm}{\ensuremath{{J_c^\pm}}}
\newcommand{\Rp}{\ensuremath{{J_r^+}}}
\newcommand{\Rm}{\ensuremath{{J_r^-}}}
\newcommand{\Rpm}{\ensuremath{{J_r^\pm}}}

\begin{document}

\title{%
  Model $I$--$V$ curves and figures of merit of underdamped deterministic Josephson ratchets
}

\author{E. Goldobin}
\author{R. Menditto}
\author{D. Koelle}
\author{R. Kleiner}
\affiliation{%
  Physikalisches Institut and Center for Quantum Science in LISA$^+$,
  Universit\"at T\"ubingen, Auf der Morgenstelle 14, D-72076 T\"ubingen, Germany
}

\date{%
  \today\ File: \textbf{\jobname.\TeX}
}

\begin{abstract}
  We propose simple models for the current-voltage characteristics of typical Josephson ratchets. We consider the case of a ratchet working against a constant applied counter force and derive analytical expressions for the key characteristics of such a ratchet: rectification curve, stopping force, input and output powers and rectification efficiency. Optimization of the ratchet performance is discussed.
\end{abstract}



\maketitle

\section{Introduction}
\label{Sec:Intro}

The discovery of Brownian motion gave birth to the idea of extracting useful work out of random motion. As Richard Feynman \etal demonstrated \cite{Feynman66}, drawing energy from equilibrium thermal fluctuations (white noise) is forbidden by the second law of thermodynamics. The extraction of work out of \emph{non-equilibrium} or time-correlated noise (colored noise) is possible using ratchet systems\cite{Magnasco:1993:Forced-Thermal-Ratchets, Juelicher:1997:MMM, Reimann:2002:BrownianMotors, Haenggi:2009:ArtBrownMotors}. As an extreme case of correlated input signal one can consider a deterministic signal with zero time average, which will be rectified by the ratchet into a dc output signal. Such \emph{deterministic ratchets} find many applications as rectifiers, sorters, \etc.
Deterministic and stochastic ratchets have been in the focus of attention during the last two decades in various implementations. In particular, Josephson systems based on the motion of Josephson vortices\cite{Falo:1999:JJA-Ratchet,Trias:2000:JJA-Ratchet,Goldobin:2001:RatchetT,Carapella:RatchetT:2001,Carapella:RatchetE:2001,Carapella:2002:JVR-HighFreq,Majer:2003:QuJVR,Ustinov:2004:BiHarmDriverRatchet,Beck:2005:RatchetE,Wang:2009:IJJ-Ratchet} or the Josephson phase \cite{Haenggi:1996:BrownRect,Zapata96,Weiss:2000:SQUID-Ratchet.E, Sterck:2002:SQUID-Ratchet, Sterck:2005:3JJ-SQUID:RockRatchet,Sterck:2009:3JJ-SQUID:StochasticRatchet} have been suggested and tested experimentally.

Josephson ratchets have some advantages over other ratchet systems: (I) directed motion results in an average dc voltage which makes ratchet operation easily accessible in experiment; (II) Josephson junctions are very fast devices, \ie, they can be operated in a broad frequency range from dc up to $100\units{GHz}$, which allows them to capture a lot of spectral energy; (III) both underdamped and overdamped systems can be investigated by proper junction design and the variation of the bath temperature.

It turns out that regardless of the underlying physics (vortex motion or Josephson phase motion) the $I$--$V$ characteristic (IVC) looks rather universal. Therefore, in this paper, we do not discuss \emph{how} such an asymmetric IVC can be obtained. Instead, we assume some specific typical IVC, parameterize it and calculate the most important figures of merit. The model presented here is an extension of a simpler model presented earlier\cite{Knufinke:2012:JVR-loaded} in two aspects. First, the present model takes into account possible hysteresis in the IVC and, therefore, allows to analyze underdamped as well as overdamped ratchets. Second, it includes two specific types of the IVC: the constant voltage step and the linear voltage branch. Within the framework of this new model we obtain rectification curves, stopping force, input and output powers and rectification efficiency.

The paper is organized as follows. In Sec.~\ref{Sec:Model} we describe the model IVCs. In Sec.~\ref{Sec:Results} the expressions for mean voltage, stopping force, input and output power and efficiency are derived. In Sec.~\ref{Sec:Discussion} we discuss the obtained results and the optimization of the ratchet. Sec.~\ref{Sec:Conclusions} concludes this work.

\section{Model}
\label{Sec:Model}

\begin{figure*}
  \centering\includegraphics{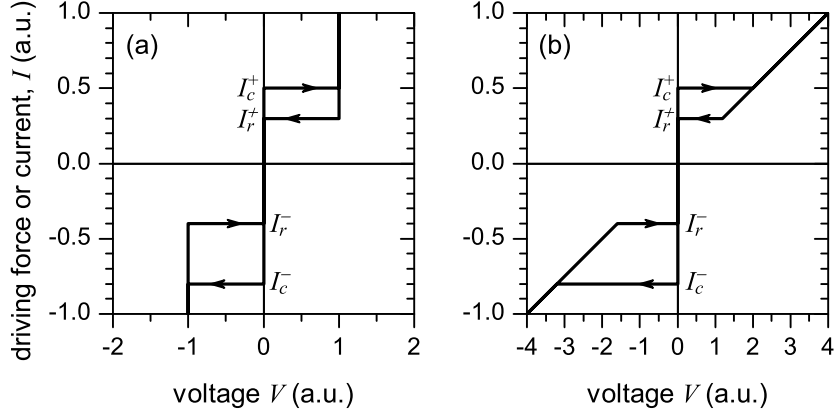}
  \caption{%
    Model IVCs for the two studied cases: (a) constant voltage step, (b) linear branch. The parameters used for plotting are: $I_c^+=0.5$, $I_c^-=0.8$, $I_r^+=0.3$, $I_r^-=0.4$,  $V_1=1$ and $R=4$.
  }
  \label{Fig:IVCs}
\end{figure*}

Since the typical frequencies of the Josephson devices are very high (from few GHz up to few THz) we consider the most simple case of the quasi-static drive $I_{ac}\sin(\omega t)$, when $\omega\ll(\omega_p,\omega_c)$. Here $\omega_p$ is the plasma frequency and $\omega_c$ is the characteristic frequency of the Josephson junction. To derive our results in the most general form from the very beginning, we assume an additional dc current (counter force) $I_{dc}$, which is applied to the ratchet trying to stop it. This allows us to study the loading capabilities of the ratchet and, thus, to calculate its output power and efficiency. The total applied driving current (force) can be written as
\begin{equation}
  I(t) = I_{dc} + I_{ac}\sin(\omega t)
  . \label{Eq:Drive.total}
\end{equation}
Taking into account the smallness of $\omega$, the rectified (mean) voltage can be obtained just by integrating a dc current-voltage characteristic $V(I)$ (most frequently measured in experiment) as
\begin{equation}
  \mean{V} =\frac{1}{T}\int_{0}^{T} V[I(t)]\,dt
  , \label{Eq:meanV.def}
\end{equation}
where $T=2\pi/\omega$ is the period of the ac drive. Note, that $V(I)$ is, in general, a hysteretic (multi-valued) function, which depends on the pre-history of the biasing.

In this work we discuss two types of IVCs shown in Fig.~\ref{Fig:IVCs}(a),(b). The first is an IVC with a constant voltage step, see Fig.~\ref{Fig:IVCs}(a), typical for relativistic motion of the phase or vortex in a Josephson device. The second is an IVC with a linear branch, see Fig.~\ref{Fig:IVCs}(b), typical for non-relativistic motion. 

The hysteresis is included in the IVC as follows. In the simplest case of only one hysteretic branch, see Figs.~\ref{Fig:IVCs}(a)--(b), we assume that when the current is increased from $I=0$, the voltage $V(I)=0$ up to $I=I_c^+$, see Fig.~\ref{Fig:IVCs} Then the voltage is given by some function ${\cal V}(I)$ specified below. However, if the current is then decreased, the voltage is given by ${\cal V}(I)$ down to the ``return current'' $I_r^+\leq I_c^+$ and $V=0$ for lower currents. If one sweeps the current $I$ in the negative direction, the corresponding critical and return currents are $-I_c^-<0$ and $-I_r^-<0$, so that $0\leq I_r^-\leq I_c^-$.

An additional bias current (counter force) $I_{dc}$, if any, shifts the origin of the ac drive from the point $I=0$ to the point $I=I_{dc}$. Alternatively, one can also treat this as an ac drive with the origin at $I=0$, but the values of $I_c^\pm$ and $I_r^\pm$ are shifted by $I_{dc}$ to the new values, $\Cpm$ and $\Rpm$ given by
\begin{eqnarray}
  \Cp = I_c^+ - I_{dc}, &\quad& \Rp = I_r^+ - I_{dc},\nonumber\\
  \Cm = I_c^- + I_{dc}, &\quad& \Rm = I_r^- + I_{dc}
  . \label{Eq:Jcr.def}
\end{eqnarray}
For the sake of simplicity, below we assume that $I_c^+<I_c^-$. Then the rectified voltage $\mean{V}\geq0$ and the counter force $I_{dc}<0$.

\section{Results}
\label{Sec:Results}

\subsection{Output (rectified) voltage}

It is convenient to calculate the \emph{average voltage} \mean{V} as a sum of average voltages over the positive and the negative periods of the drive, \ie,
\begin{equation}
  \mean{V} = \mean{V}_+ + \mean{V}_-
  . \label{Eq:meanV}
\end{equation}
where
\begin{equation}
  \mean{V}_\pm(I_{ac}) = \begin{cases}
    0,                                  &\text{if }I_{ac}<\Cpm\\
    \mean{\cal V}_\pm(I_{ac}),&\text{if }I_{ac}>\Cpm
  \end{cases}
  . \label{Eq:meanVpm.gen}
\end{equation}
The functions $\mean{\cal V}_\pm(I_{ac})$ will be calculated below for a particular model as follows.

\begin{align}
  \mean{\cal V}_{\pm} &\stackrel{\eqref{Eq:meanV.def}}{=}
  \frac{\omega}{2\pi}\int^{t_2^\pm}_{t_1^\pm} {\cal V}[I_{dc}+I_{ac}\sin(\omega t)] dt\nonumber\\
  &=\pm\frac{1}{2\pi} \int^{\phi_2^\pm}_{\phi_1^\pm} {\cal V}[I_{dc}+I_{ac}\sin(\phi)] d\phi \nonumber\\
  , \label{Eq:1br:Vpm.pre}
\end{align}
where $\phi_1^\pm$ and $\phi_2^\pm$ define the phases ($t_1^\pm$ and $t_2^\pm$ define times) when the system switches to and from the voltage state, \ie, the instant value of current $I(t)$ exceeds $I_c^\pm$ or falls below $I_r^\pm$, \ie, $I_{ac}\sin(\omega t)$ exceeds $\Cpm$ or falls below $\Rpm$. They are given by
\begin{subequations}
  \begin{eqnarray}
      \omega t_1^+\equiv\phi_1^+ &=& \phantom{2\pi}+\arcsin(\Cp/I_{ac}),\\
      \omega t_2^+\equiv\phi_2^+ &=& \phantom{2}\pi-\arcsin(\Rp/I_{ac}),\\
      \omega t_1^-\equiv\phi_1^- &=& \phantom{2}\pi+\arcsin(\Cm/I_{ac}),\\
      \omega t_2^-\equiv\phi_2^- &=& 2\pi-\arcsin(\Rm/I_{ac}).
  \end{eqnarray}
  \label{Eq:phi12}
\end{subequations}

\textbf{Constant voltage model.} For a constant voltage model\cite{Knufinke:2012:JVR-loaded}, see Fig.~\ref{Fig:IVCs}(a),
\begin{equation}
  {\cal V}(I) = V_1 \sgn(I)
  , \label{Eq:IVC:V=const}
\end{equation}
relevant for step-like behavior of the IVC, see Fig.~\ref{Fig:IVCs}(a). Here, $V_1$ is the voltage of the step.

Substituting \eqref{Eq:IVC:V=const} into Eq.~\eqref{Eq:1br:Vpm.pre} we obtain
\begin{align}
  \mean{\cal V}_{\pm} = \pm\frac{V_1}{2\pi} \left( \phi_2^\pm-\phi_1^\pm \right)
  , \label{Eq:const:Vpm.pre}
\end{align}
and using Eqs.~\eqref{Eq:phi12} we obtain the final explicit expression
\begin{align}
  \mean{\cal V}_\pm
  &= \pm\frac{V_1}{2\pi} \left[
    \arccos\left(\frac{\Rpm}{I_{ac}}\right)+\arccos\left(\frac{\Cpm}{I_{ac}}\right)
  \right].
  \label{Eq:const:Vpm}
\end{align}

\textbf{Linear voltage model.} For the linear voltage model, see Fig.~\ref{Fig:IVCs}(b),
\begin{equation}
  {\cal V}(I) = R_n I
  , \label{Eq:IVC:V=IR}
\end{equation}
relevant for IVCs obtained from the resistively and capacitively shunted junction (RCSJ) model with resistance $R_n$, Fig.~\ref{Fig:IVCs}(b).

Substituting \eqref{Eq:IVC:V=IR} into Eq.~\eqref{Eq:1br:Vpm.pre} we obtain
\begin{align}
  \mean{\cal V}_{\pm}
  = \frac{R_n}{2\pi} \left\{
    I_{dc}(\phi_2^\pm - \phi_1^\pm) + I_{ac} \left[ \cos(\phi_1^\pm)-\cos(\phi_2^\pm) \right]
  \right\}
  . \label{Eq:lin:Vpm.pre}
\end{align}
Using Eqs.~\eqref{Eq:phi12} and taking into account that $\cos(\phi_1^\pm)\gtrless 0$, while $\cos(\phi_2^\pm)\lessgtr 0$ we obtain
\begin{widetext}
\begin{equation}
  \mean{\cal V}_{\pm} =
  \pm\frac{R_n}{2\pi} \left\{
    I_{dc} \left[ \arccos\left(\frac{\Rpm}{I_{ac}}\right)+\arccos\left(\frac{\Cpm}{I_{ac}}\right) \right]+
    I_{ac} \left[ \sqrt{1-\left(\frac{\Cpm}{I_{ac}}\right)^2}+\sqrt{1-\left(\frac{\Rpm}{I_{ac}}\right)^2} \right]
  \right\}
  . \label{Eq:lin:Vpm.loaded}
\end{equation}
\end{widetext}
Note, that the first term in Eqs.~\eqref{Eq:lin:Vpm.pre} and \eqref{Eq:lin:Vpm.loaded} vanishes for $I_{dc}=0$ (idle ratchet).

\begin{figure*}
  \centering\includegraphics{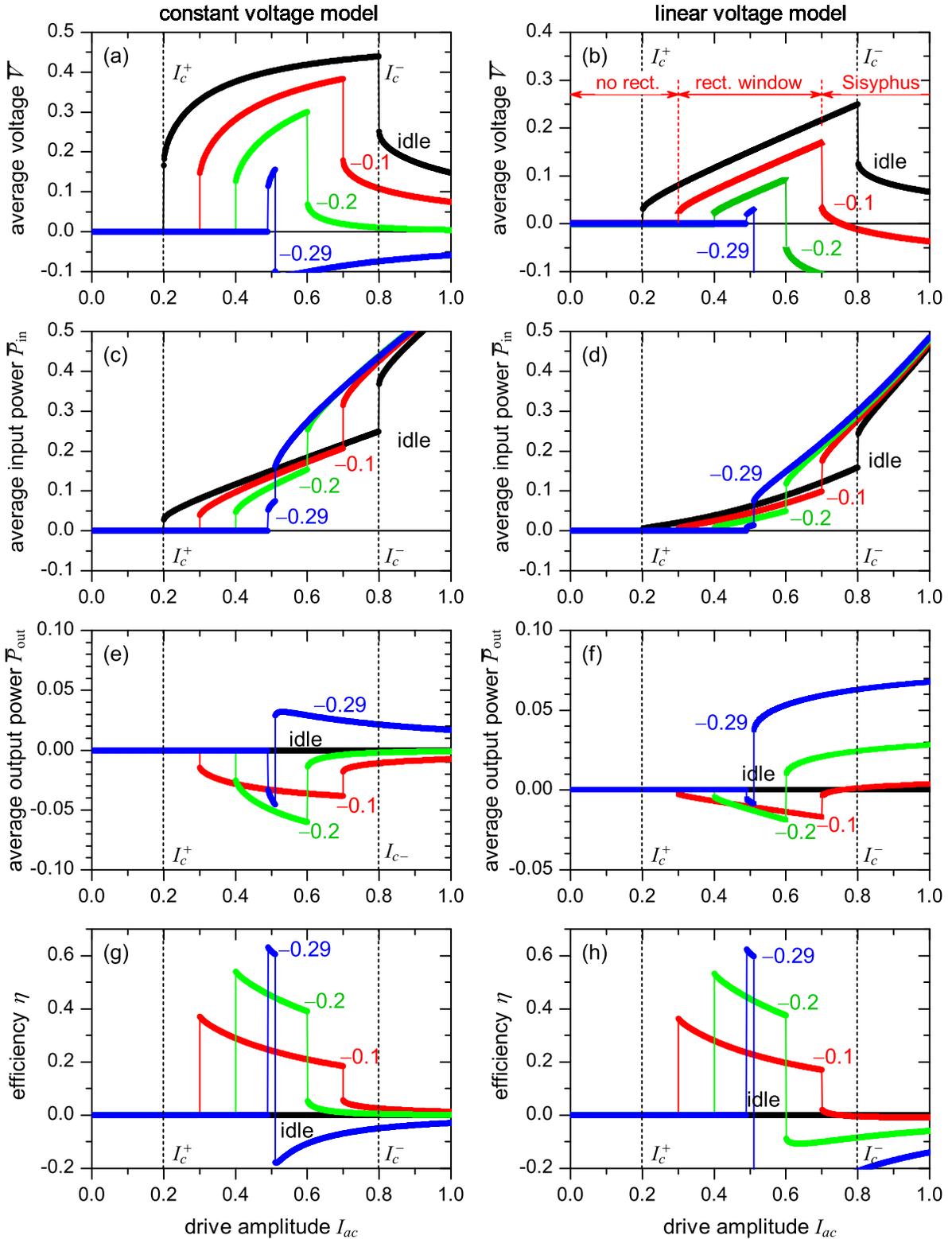}
  \caption{%
    A set of rectification curves $\mean{V}(I_{ac})$ (a,b), inpit power $P_\mathrm{in}(I_{ac})$ (c,d), output power $P_\mathrm{out}(I_{ac})$ (e,f) and efficiency $\eta(I_{ac})$ calculated for different values of $I_{dc}=0,\,-0.1,\,-0.2,\,-0.29$.
    The left column of plots, \ie, (a,c,e,g), is calculated using the constant voltage model with $V_1=1$. The right column, \ie, (b,d,f,h), is calculated using the linear voltage model with $R_n=1$. The other parameters are $I_c^+=0.2$, $I_c^-=0.8$, $I_r^+=0.1$, $I_r^-=0.3$. In (b) the most important regions are marked for the case of a ratchet loaded by $I_{dc}=-0.10$.
  }
  \label{Fig:Vav(Iac)@Idc}
\end{figure*}

\subsection{Input, output power and efficiency}

Without applied dc bias (counter force) the ratchet rectifies, but it is idle, \ie, it does not produce any output power. Therefore, for the analysis of power input, output and efficiency we consider the case $I_{dc}\neq0$, \ie, the ratchet should work against an applied dc current (counter force) $I_{dc}$.

A set of rectification curves for several values of $I_{dc}\leq0$ is shown in Fig.~\ref{Fig:Vav(Iac)@Idc}(a),(b). The general behavior of $\mean{V}(I_{ac})$ follows from Eqs.~\eqref{Eq:meanV} and \eqref{Eq:meanVpm.gen}. For small values of the drive $I_{ac}<\Cp$ the value of $\mean{V}=0$. Then, for $\Cp<I_{ac}<\Cm$, we get a strong rectification because $\mean{V}_+>0$ while $\mean{V}_-=0$. Finally, at $I_{ac}>\Cm$, both $\mean{V}_+>0$ and $\mean{V}_-<0$ almost cancel each other.

The no-rectification regime at $I_{ac}<\Cp$ we shall call a pinning regime. We define a \emph{rectification window} as a range of $I_{ac}$ where rectified voltage is large, \ie, $\Cp<I_{ac}<\Cm$. The region $I_{ac}>\Cm$ we shall call a ``Sisyphus'' regime since the system makes a lot of useless back and force motion, dissipating a lot of power, but not producing any appreciable mean output. We do not pay much attention to the Sisyphus region since it is not interesting for applications.

\begin{figure}[!htb]
  \centering\includegraphics{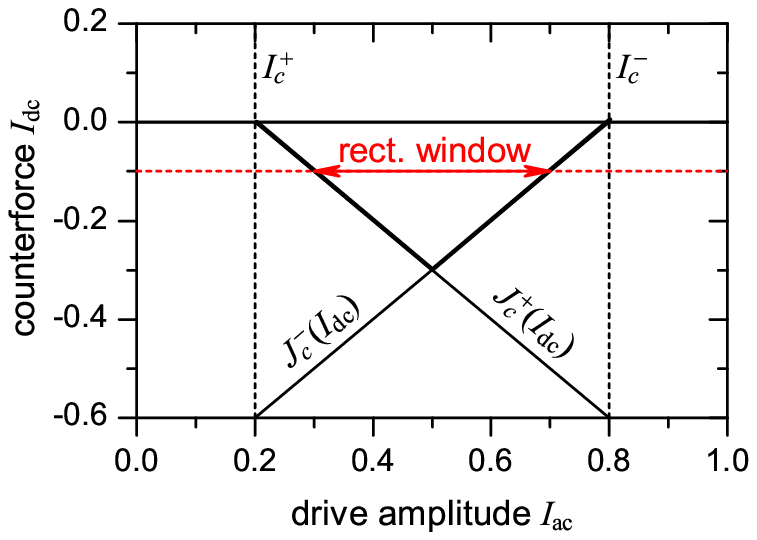}
  \caption{%
    The size of rectification window at given $I_{dc}$. As an example the arrow shows the size of rectification window at $I_{dc}=-0.1$. Simultaneously this plot shows the stopping force $I_\mathrm{stop}$ at given drive amplitude $I_{ac}$.
  }
  \label{Fig:StopForce}
\end{figure}

From Fig.~\ref{Fig:Vav(Iac)@Idc} we see that the rectification window shrinks as the absolute value of the counter force $I_{dc}<0$ increases. Let us plot the beginning and the end of the rectification window, \ie, $\Cpm(I_{dc})$ given by Eq.~\eqref{Eq:Jcr.def}, see Fig.~\ref{Fig:StopForce}. The window shrinks linearly with $I_{dc}$. The two lines $\Cp(I_{dc})$ and $\Cm(I_{dc})$ cross (and the rectification window closes) at $I_{dc}=(I_c^+ - I_c^-)/2$. Just before closing, the rectification takes place only at the single value of $I_{ac}=(I_c^+ + I_c^-)/2$, \ie, just in the middle of the idle ($I_{dc}=0$) rectification window.

The \emph{stopping force} $I_\mathrm{stop}$ is defined as the dc counter force $I_{dc}$ that one has to apply at fixed $I_{ac}$ to stop the ratchet. Since we are ignoring the Sisyphus region, the stopping force is basically defined by the motion of boundaries of the rectification window, see Fig.~\ref{Fig:StopForce}, \ie it is determined only by $I_c^\pm$ and is independent on exact details of ${\cal V}(I)$, provided it is point-symmetric, \ie, ${\cal V}(I)=-{\cal V}(-I)$. The stopping force is given by\footnote{Our previous paper \cite{Knufinke:2012:JVR-loaded} contains erroneous expression for $I_\mathrm{stop}$ for $I_{ac}>(I_c^+ + I_c^-)/2$},
\begin{equation}
  I_\mathrm{stop} =
  \begin{cases}
    0,              &\text{for } I_{ac}<I_c^+;\\
    I_c^+ - I_{ac}, &\text{for } I_c^+ < I_{ac} < (I_c^+ + I_c^-)/2;\\
    I_{ac}- I_c^- , &\text{for } (I_c^+ + I_c^-)/2<I_{ac} < I_c^-;\\
    0               &\text{for } I_c^+<I_{ac}\text{ (Sisyphus)}
  \end{cases}
  . \label{Eq:Istop}
\end{equation}

The mean \emph{input ac power} is given by
\begin{eqnarray}
  \mean{P}_\mathrm{in}
  &=& \frac1{2\pi} \int_{0}^{2\pi} V[I_{dc}+I_{ac}\sin(\phi)] I_{ac}\sin(\phi)\,d\phi
  \nonumber\\
  &=& \mean{P}_\mathrm{in}^+ + \mean{P}_\mathrm{in}^-
  , \label{Eq:def:Pin}
\end{eqnarray}
where the power during the positive and negative semi-periods are given by
\begin{equation}
  \mean{P}_\mathrm{in}^\pm =
  \begin{cases}
    0,                            & \text{ for } I_{ac}<\Cpm\\
    \mean{\cal P}_\mathrm{in}^\pm & \text{ for } I_{ac}>\Cpm.
  \end{cases}
  \label{Eq:Pin:pm}
\end{equation}
Here
\begin{equation}
  \mean{\cal P}_\mathrm{in}^\pm
  = \frac1{2\pi} \int_{\phi_1^\pm}^{\phi_2^\pm} V[I_{dc}+I_{ac}\sin(\phi)] I_{ac}\sin(\phi)\,d\phi
  . \label{Eq:calPin:pm}
\end{equation}

The mean \emph{output dc power} for constant dc bias current $I_{dc}$ is given by
\begin{equation}
  \mean{P}_\mathrm{out} = \frac1{2\pi} \int_{0}^{2\pi} V(I_{dc}+I_{ac}\sin\phi) I_{dc}\,d\phi = \mean{V}I_{dc}
  . \label{Eq:P_out}
\end{equation}
Note that for $I_c^+ < I_c^-$ the rectified voltage $\mean{V}\geq 0$, so that we apply a stopping current $I_{dc}<0$. Thus, $\mean{P}_\mathrm{out}<0$, which indicates that the power is not consumed but rather given out to the dc load (dc current source). Eq.~\eqref{Eq:P_out} says that $\mean{P}_\mathrm{out}(I_{dc})$ does not require any separate calculations and can be obtained from $\mean{V}(I_{ac})$ in a trivial way.

The \emph{efficiency} is given by
\begin{equation}
  \eta = - \mean{P}_\mathrm{out}/\mean{P}_\mathrm{in}
  . \label{Eq:Eff}
\end{equation}

\textbf{Constant voltage model.} For model \eqref{Eq:IVC:V=const}, after integration of Eq.~\eqref{Eq:calPin:pm} we have
\begin{equation}
  \mean{\cal P}_\mathrm{in}^\pm =
  \frac{V_1 I_{ac}}{2\pi} \left[
    \sqrt{1-\left(\frac{\Cpm}{I_{ac}}\right)^2}+ \sqrt{1-\left(\frac{\Rpm}{I_{ac}}\right)^2}
  \right]
  \label{Eq:const:calP_in}
\end{equation}
The plots $P_\mathrm{in}(I_{ac})$ for different values $I_{dc}$ are shown in Fig.~\ref{Fig:Vav(Iac)@Idc}(c), while $P_\mathrm{out}(I_{ac})$ is shown in Fig.~\ref{Fig:Vav(Iac)@Idc}(e).

The efficiency $\eta(I_{ac})$ can be calculated explicitly using Eq.~\eqref{Eq:Eff}. We do not show this bulky expression here, however, we plot the result in Fig.~\ref{Fig:Vav(Iac)@Idc}(g). The efficiency has a maximum just at the beginning of the rectification window, \ie, at $I_{ac}=\min(\Cp,\Cm)$. Assuming that $\Cp<\Cm$, which is always the case for the still open rectification window, the maximum efficiency is reached at $I_{ac}=\Cp$ and is given by
\begin{equation}
  \eta_\mathrm{max}
  = \frac{-I_{dc}}{2}\frac{\arccos\left( \frac{\Rp}{\Cp} \right)}{\sqrt{\Cp^2 - \Rp^2}}
  . \label{Eq:const:Eff.max}
\end{equation}

\textit{No hysteresis.} In the case of no hysteresis ($I_r^+ \to I_c^+$, \ie, $\Rp \to \Cp$) we recover the previous result\cite{Knufinke:2012:JVR-loaded}, namely $\eta\to -I_{dc}/\Cp=-I_{dc}/(I_c^+-I_{dc})$. This is a monotonically increasing function of $-I_{dc}$. However, $-I_{dc}$ cannot be made arbitrary large. At $I_{dc}\to(I_c^+ - I_c^-)/2$ the rectification window is about to close, but efficiency approaches its ultimate value given by
\begin{equation}
  \eta_\mathrm{ult}\to \frac{I_c^- - I_c^+}{I_c^- + I_c^+}
  . \label{Eq:const:Eff.ult.nohyst}
\end{equation}

\textit{Maximum hysteresis.} In this case ($I_r^+ = 0$) we obtain the ultimate rectification at $I_{dc}\to(I_c^+ - I_c^-)/2$ (just before rectification window closes) given by
\begin{equation}
  \eta_\mathrm{ult}\to \frac{I_c^- - I_c^+}{2\sqrt{I_c^- I_c^+}}
  \arccos\left( \frac{I_c^- - I_c^+}{I_c^- + I_c^+} \right)
  . \label{Eq:const:Eff.ult.mhyst}
\end{equation}

Let us make several notes. First, when we speak about efficiency (and discuss effects of hysteresis), we focus on the rectification window only. There, only the hysteresis of the positive branch $I_r^+$ matters because the negative branch is not participating yet. Thus, the value of $I_r^-$ is irrelevant, while the value of $I_c^-$ only affects the upper edge of the rectification window. Second, to make the ratchet more efficient one has to design it with $I_c^+ \to 0$, while keeping $I_c^-$ constant. In this limit the hysteresis of the positive part of the IVC (the value of $I_r^+$) becomes irrelevant too since $I_r^+ < I_c^+$, \ie, $I_r^+=0$.

\textbf{Linear voltage model.} For model \eqref{Eq:IVC:V=IR}, after integration of Eq.~\eqref{Eq:calPin:pm} we have

\begin{widetext}
\begin{equation}
  \mean{\cal P}_\mathrm{in}^\pm =
  \frac{I_{ac}R_n}{4\pi}\left\{
    (\Cpm \pm 2I_{dc})\sqrt{1-\frac{\Cpm^2}{I_{ac}^2}} +
    (\Rpm \pm 2I_{dc})\sqrt{1-\frac{\Rpm^2}{I_{ac}^2}}
    + I_{ac} \left[ \arccos\left( \frac{\Cpm}{I_{ac}} \right) + \arccos\left( \frac{\Rpm}{I_{ac}} \right)\right]
  \right\}
  . \label{Eq:lin:calP_in}
\end{equation}
\end{widetext}
The efficiency $\eta(I_{ac})$ can be calculated explicitly using Eq.~\eqref{Eq:Eff}. We do not show this bulky expression here, however, we plot the result in Fig.~\ref{Fig:StopForce}(h). The efficiency has a maximum at the beginning of the rectification window, \ie, at $I_{ac}=\min(\Cp,\Cm)$. Assuming that $\Cp<\Cm$, which is always the case for the still open rectification window, the maximum efficiency is reached at $I_{ac}=\Cp$ and is given by
\begin{equation}
  \eta_\mathrm{max} = \frac{-2I_{dc}\left[
    I_{dc} \arccos\left( \frac{\Rp}{\Cp} \right) + \sqrt{\Cp^2-\Rp^2}
  \right]}
  {\Cp^2 \arccos\left( \frac{\Rp}{\Cp} \right) + \sqrt{\Cp^2 - \Rp^2}(\Rp+2I_{dc})}
  . \label{Eq:lin:Eff.max}
\end{equation}

\textit{No hysteresis.} Note that in the limit of no hysteresis ($\Rp \to \Cp$) we again recover the previous result\cite{Knufinke:2012:JVR-loaded}, see Eq.~\eqref{Eq:const:Eff.ult.nohyst}. This is independent on our model (linear branch or constant voltage branch) as at the beginning of rectification window the branch is not really traced yet (only its first point).

\textit{Maximum hysteresis.} In this case ($I_r^+ = 0$) we obtain the ultimate rectification at $I_{dc}\to(I_c^+ - I_c^-)/2$ (just before rectification window closes) given by
\begin{equation}
  \eta_\mathrm{ult} = \frac{-2(I_c^- - I_c^+)^2\arccos\left( \frac{I_c^- - I_c^+}{I_c^- + I_c^+} \right) + 4(I_c^- - I_c^+)\sqrt{I_c^+ I_c^-}}
  {(I_c^+ + I_c^-)^2 \arccos\left( \frac{I_c^- - I_c^+}{I_c^- + I_c^+} \right) + 2(I_c^+ - I_c^-)\sqrt{I_c^+ I_c^-}}
  . \label{Eq:lin:Eff.ult.mhyst}
\end{equation}

\section{Discussion.}
\label{Sec:Discussion}

The plots of $\mean{P}_\mathrm{in}(I_{dc})$ and $\mean{P}_\mathrm{out}(I_\mathrm{dc})$ as well as $\eta(I_{dc})$ for both models are shown in Fig.~\ref{Fig:Vav(Iac)@Idc}(c)--(h). From Fig.~\ref{Fig:Vav(Iac)@Idc}(c),(d) we see that the input power is zero for $I_{ac}<\Cp$, increases inside the rectification window, and increases even further in the Sisyphus regime. As we apply the counter force $I_{dc}$, the input power within the rectification window decreases slightly.

The behavior of $\mean{P}_\mathrm{out}(I_{ac})$ is more complicated. First of all, for the case of the idle/unloaded ratchet ($I_{dc}=0$) $\mean{P}_\mathrm{out}(I_{ac}) \equiv 0$ for any $I_{ac}$. For the loaded ratchet ($I_{dc}<0$), the power $\mean{P}_\mathrm{out}(I_{ac})=0$ in the pinning regime, and $\mean{P}_\mathrm{out}(I_{ac})$ has its maximum value inside the rectification window. In the Sisyphus regime the power $\mean{P}_\mathrm{out}(I_{ac})$ becomes small or even changes sign (power consumption instead of power generation). Interestingly, the maximum value $\mean{P}_\mathrm{out}(I_{ac})$ is reached for some $I_{dc}$ in the middle of the interval $(I_c^+ - I_c^-)/2 \ldots 0$.

The efficiency $\eta(I_{ac})$ has a more clear behavior. It has a maximum in the beginning of the rectification window and grows as the load, \ie, $|I_{dc}|$, increases.

From a practical point of view we would like to choose the parameters $I_c^\pm$, $I_r^\pm$, $V_1$ or $R_n$ to optimize the following figures of merit of our Josephson ratchet. (a) The \emph{rectification window} should be made as wide as possible and it should start at the lowest possible $I_{ac}$. The stopping force/current $I_\mathrm{stop}$ is directly related to the rectification window size, so that a large window automatically leads to a high $|I_\mathrm{stop}|$. (b) The output (rectified) voltage $\mean{V}$ should be made as high as possible. (c) The output power $P_\mathrm{out}$ should be made as large as possible. (d) The efficiency $\eta$ should be made as large as possible.

The parameters $V_1$ and $R_n$ have a direct effect on $\mean{V}$ and $\mean{P}_\mathrm{out}$ and have no effect on the rectification window and efficiency. Therefore to increase $\mean{V}\propto (V_1,R_n)$ and $\mean{P}_\mathrm{out}\propto (V_1,R_n)$ one should increase $V_1$ or $R_n$.

To enlarge the rectification window one has to make $I_c^+$ and $I_c^-$ as different as possible (maximum possible asymmetry). In an ideal case, one would like to have $I_c^+ \to 0$ (which automatically means $I_r^+\to0$). The values of $I_r^\pm$ are not relevant for the rectification window size. Simultaneously, a large rectification window leads to large values of the stopping force up to $(I_c^+ - I_c^-)/2$. The efficiency has its maximum value in the beginning of the rectification window and also grows with increasing load current $|I_{dc}|$. However, with increasing load the rectification window shrinks. At the end one has to find a compromise between efficiency and rectification window size for a particular application. If the value of $I_c^+\neq0$ the value of $I_r^+$ does not have a major effect on the rectification curves -- smaller $I_r^+$ improves the figures of merit such as $\mean{V}$, $\mean{P}_\mathrm{out}$ and $\eta$ somewhat. The value of $I_r^-$ is relevant only in the Sisyphus region, which is not interesting for applications.

\section{Conclusions}
\label{Sec:Conclusions}

We have suggested two simple models of a Josephson ratchet --- the linear voltage branch and the constant voltage branch --- based on few experimentally relevant parameters. We have derived analytical expressions for the mean voltage $\mean{V}$, the rectification window size, the stopping force $I_\mathrm{stop}$, the input power $\mean{P}_\mathrm{in}$, the output power $\mean{P}_\mathrm{out}$ and efficiency $\eta$. We have demonstrated the performance of the ratchet for some typical set of parameters and discussed the optimization of different figures of merit. This results should be useful for designing the next generation of Josephson ratchets as well as for fitting already obtained results.

\acknowledgments

R.M. gratefully acknowledges support by the Carl Zeiss Stiftung. This work was supported by the Deutsche Forschungsgemeinschaft (DFG) via Project No. GO-1106/5, via project A5 within SFB/TRR-21, and by the EU-FP6-COST action MP1201.

\clearpage
\bibliography{SF,pi,software,ratch,LJJ}

\end{document}